\documentclass[preprints,article,accept,moreauthors,pdftex]{Definitions/mdpi}

\firstpage{1}
\pubvolume{1}
\issuenum{1}
\articlenumber{0}
\pubyear{2022}
\copyrightyear{2022}
\datereceived{}
\dateaccepted{}
\datepublished{}
\hreflink{https://doi.org/} 
\pdfoutput=1

\usepackage{textgreek}

\Title{Deep learning study of an electromagnetic calorimeter}

\TitleCitation{Deep learning study on an electromagnetic calorimeter}

\Author{Elihu Sela $^{1}$, Shan Huang $^{1}$ and David Horn $^{1}$*}

\AuthorNames{Elihu Sela, Shan Huang and David Horn}
\AuthorCitation{Sela, E.; Huang, S.; Horn, D.}

\address[1]{%
$^{1}$ \quad Tel Aviv University, Tel Aviv-Yafo 6997801, Israel}

\corres{Correspondence: horn@tau.ac.il}

\abstract{The accurate and precise extraction of information from a modern particle physics detector, such as an electromagnetic calorimeter, may be complicated and challenging.
In order to overcome the difficulties we propose processing the detector output using the deep-learning methodology.
Our algorithmic approach makes use of a known network architecture, which is being modified to fit the problems at hand.
The results are of high quality (biases of order 2\%) and, moreover, indicate that most of the information may be derived from only a fraction of the detector.
We conclude that such an analysis helps us understanding the essential mechanism of the detector and should be performed as a part of its designing procedure.}

\begin{document}

\section{Introduction}
In particle physics, calorimetry is a method of shower detection to identify the incident particle's energy and position coordinates \cite{hep-perkins}.
The electromagnetic calorimeter (ECAL) is a detector where the incident particles are showered and absorbed via the electromagnetic interaction, such as bremsstrahlung and pair creation.
To obtain a snapshot of the shower, multiple layers of sensitive materials are inserted in the ECAL where the shower's energy is deposited and the image of shower energy distribution is built.
Via Monte--Carlo simulation \cite{Geant4}, the procedures of showering and energy deposition are studied. This forms an integral part of the design of a detector system.
Given that the simulation concerns a multitude of active elements, it requires further interpretations, such as retrieving answers from the detector which agree with the simulation.

Here we present an example of the design for an ECAL detector in the proposed \textit{Laser und XFEL Experiment} (LUXE) \cite{LUXE_CDR-2021}.
The ECAL is presented as a 3D structure of $(110\times 11\times 20 =)$ 24,200 elements of detector pads.
This detector is exposed to a beam of incoming particles, specifically, positrons produced by a laser beam colliding with electrons.
During the experiment, the positrons will be generated periodically. Each batch of the positrons, in chronological order, is called an ``event''.
The physical results of the collision will change along with laser beam's properties, which in this case is qualified by a single parameter: the focal radius of the laser beam. In this article, the samples with radii $\rho=3,5,8$ \textmu m are investigated.
The \textbf{energy distribution} and \textbf{multiplicity} (defined as the number of particles that hit the detector) of the positron beam vary not only along with the laser properties, but also due to statistical and quantum fluctuations.

The question to be answered is how can one find an algorithm that translates the readings of the 24-thousand quasi-continuous variables to the aforementioned physical properties, and covers the different scenarios where the incident beam includes large variation in the number of positrons.
In principle, the readings of each event on the detector becomes a data point in the space with 24K dimensions. Even an event of thousands of positrons occupies only a small volume in this large space.
The challenge of working within such a data-space may be answered by a deep neural network. The field of deep learning has developed in the past decade to encounter such tasks.

\begin{figure}[H]
\includegraphics[width=10.5 cm]{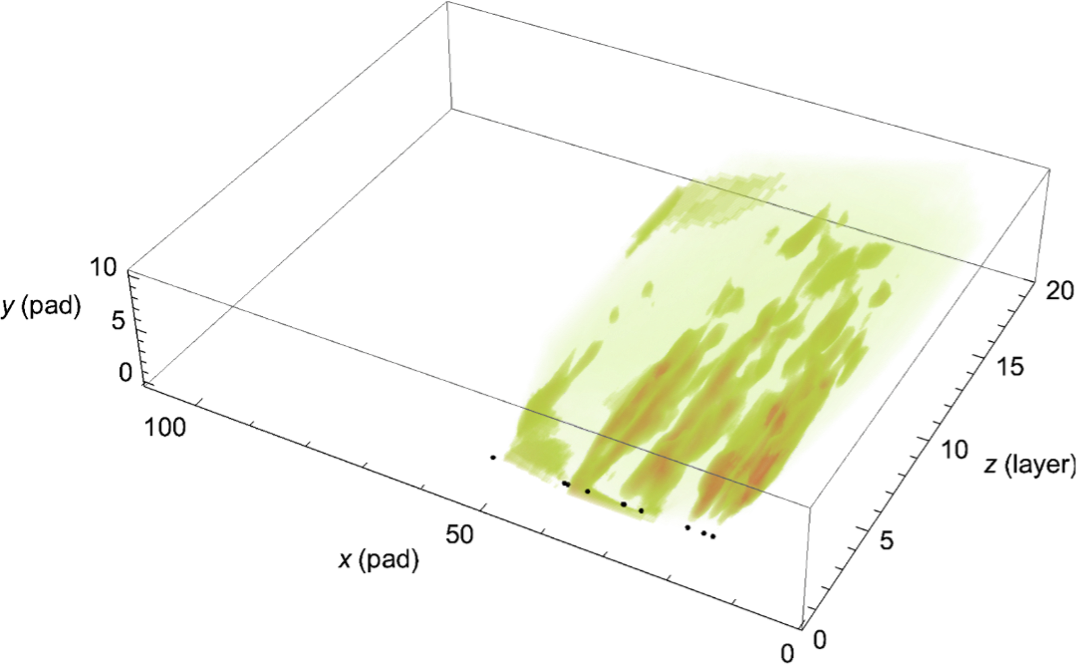}
\caption{Simulation of the shower's energy deposition expected to occur within the detector upon being hit by 10 positrons (the black dots). In reality, a ``pad'' in $x$-$y$ plane take an area of $5\times 5$ mm$^{2}$, and the distance between two neighbor ``layers'' along $z$ direction is 4.5 mm.\label{fig_1}}
\end{figure}

In this article, we demonstrate the results obtained from a convolutional neural network (CNN) based on the successful residual network ResNet10 \cite{He_ResNet-2016}.
We apply this tool to a detector image of 110-by-11-by-20 elements in a 3D Euclidean space with $(x,y,z)$ coordinates.
Fig.~\ref{fig_1} sketches an expected shower structure within the detector.

\unskip

\section{Methods}

The original ResNet has excelled in analysis of colored images \cite{He_ResNet-2016}. The input layer of the original ResNet had 2 space dimensions and 1 color dimension containing the images in 3 color filters.
For the ECAL, all 3 spatial axes start out as physical distances specifying locations in the detector.
The architecture of our CNN is demonstrated in Table \ref{tab_1}, using a customary ResNet specifications \cite{Feng-2017}.
A three-dimensional convolution leads to the first layer of a reduced $x$ dimension.
Following it are four ResNet blocks containing 2 or 3 layers, which are interconnected via 3D convolutions, with varying numbers of feature map dimensions.
The final layer is a linear vector of size 1024 connected to the output.
The total 14.4 million network parameters are trained by stochastic gradient descent.
The output can be a moment of the energy distribution of the shower (Sec. \ref{sec_3}), the discrete energy spectrum consisting of an array of histogram bins (Sec. \ref{sec_4}), or the multiplicity $N$ (Sec. \ref{sec_5}).
The open-source code employed for our calculations is provided in GitHub \cite{Sela_source_code-2021}.

\begin{table}[H]
\caption{The ResNet10 overall structure, using the customary ResNet specifications \cite{Feng-2017}.\label{tab_1}}
\newcolumntype{C}{>{\centering\arraybackslash}X}
\begin{tabularx}{\textwidth}{lCCC}
\toprule
\textbf{Type} & \textbf{3D structure} & \textbf{Number of layers} & \textbf{Number of features}\\
\midrule
Input	& [110, 11, 21]	&	&\\
Layer 1	& [31, 11, 21]	& 1	& 64\\
Block 1	& [16, 6, 11]	& 2	& 64\\
Block 2	& [8, 3, 6]		& 3	& 128\\
Block 3	& [4, 2, 3]		& 3	& 256\\
Block 4	& [2, 1, 2]		& 3	& 512\\
Linear	&				& 1	&\\
Output	&				&	&\\
\bottomrule
\end{tabularx}
\end{table}
\unskip

\section{Positron energy distribution}
An ideal detector output is a list of energies for the particles generated in each event.
But for multiple showers covering each other, a more feasible result would be the energy distribution function $f(E)$ defined as in
\begin{linenomath}
\begin{equation} \label{eq_1}
    n|_{E_1}^{E_2} = \int_{E_1}^{E_2}{f(E)\,\mathrm{d}E}.
\end{equation}
\end{linenomath}
The number of positrons with energies in between $E_1$ and $E_2$ is denoted by $n|_{E_1}^{E_2}$.
The average energy distribution function in the experimental setup which we study is characterized by the focal radius $\rho$ of the laser which is being employed .
Following in this section, two aspects of $f(E)$ will be studied using our CNN methodology.

\subsection{Moments of the distribution}
\label{sec_3}

We can characterize the distribution by its moments
\begin{linenomath}
\begin{equation}
    M_i = \left< E^i \right> = \int{E^i f(E)\,\mathrm{d}E}.
\end{equation}
\end{linenomath}
Specifically,
\begin{linenomath}
\begin{equation}
    M_1 = \text{mean},\;\;\;\text{and}\;\;\;
    M_2-M_1^2 = \text{variance}=\sigma^2.
\end{equation}
\end{linenomath}
In each event (labelled $a$), a finite number of positrons will be generated. This number varies between events.
The discrete definition of moments is used and weighted by the multiplicity of each event $n_a$.
We denote the energy moments of each such configuration by $m_{a,i}$, and then reconstruct the overall moments through
\begin{linenomath}
\begin{equation} \label{eq_4}
    M_i = \sum_{a} {\frac{n_a m_{a,i}}{N}} \;\;\;\text{where}\,
        N = \sum_{a} {n_a}.
\end{equation}
\end{linenomath}

\begin{figure}[H]
\includegraphics[width=7 cm]{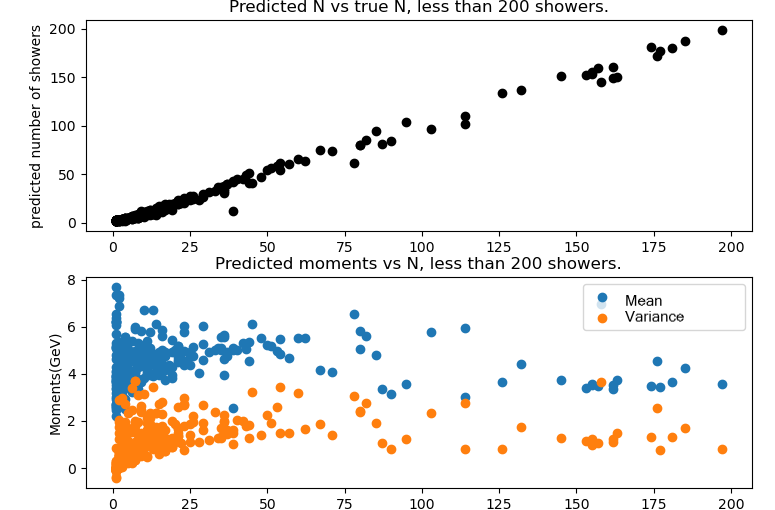}
\includegraphics[width=7 cm]{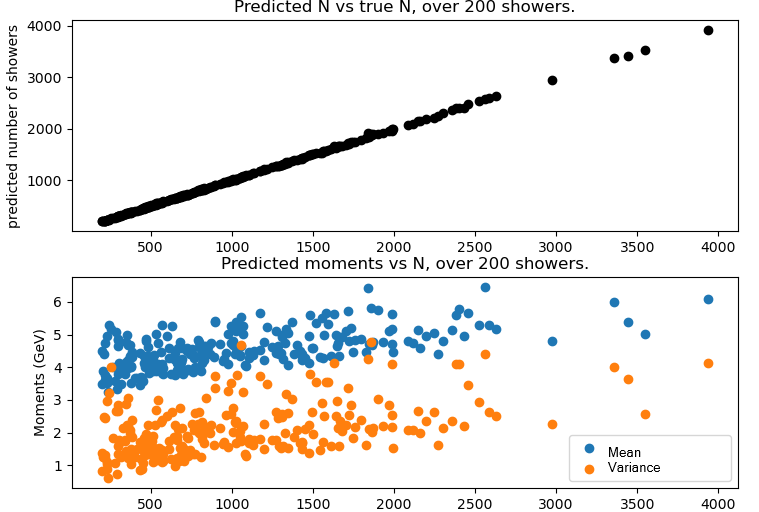}
\caption{CNN predicted values of the mean and variance, for each event $a$. The events are ordered according to their number of positrons $n_a$. The left and right frames refer to the smaller ($<200$) and larger ($\ge 200$) numbers of positrons. The predictions of the CNN agree with the values of the Monte--Carlo (MC) data. These results are used to evaluate the total moments of the energy distributions displayed in Table \ref{tab_2}.\label{fig_2}}
\end{figure}

We obtain for each event both the multiplicity $n_a$ and the various moments $m_{a,i}$.
Characteristically, we are interested in the first few moments.
The CNN is trained by a randomly chosen 75\% of the dataset with one event at a time, and tested by the remained 25\%.
Different CNNs are used to train for the multiplicity $n_a$ of the event, as registered by ECAL, and each one of its energy moments $m_{a,i}$.
The results are displayed in Fig.~\ref{fig_2}.

\unskip

\begin{table}[H]
\caption{Moments of the energy distribution of the test set of the $\rho=3$ \textmu m source, summing over all events. The true values are determined from the MC data, and the reconstructed values arise from the CNN evaluation of the ECAL analysis of individual events, as described above. \label{tab_2}}
\newcolumntype{C}{>{\centering\arraybackslash}X}
\begin{tabularx}{\textwidth}{C|CCC}
\toprule
\textbf{Moments}	& \textbf{True Results}		& \textbf{Reconstructed}		& \textbf{Bias}\\
\midrule
$M_1$		& 4.65		& 4.71		& 1.2\%\\
$M_2$		& 24.1		& 24.6		& 2.3\%\\
$M_3$		& 137.4		& 142.8		& 3.9\%\\
\bottomrule
\end{tabularx}
\end{table}
\unskip

\subsection{Discrete positron energy spectrum}
\label{sec_4}

Apart from the characteristic parameters of the distribution, another straightforward method is often used by breaking apart the continuous distribution function into a discrete histogram like Eq. (\ref{eq_1}).
Here we train and test the data on an energy spectrum with 20 bins. The energy ranges of these bins are predetermined as $\left[E_i, E_{i+1}\right)$ and

\begin{figure}[H]
\includegraphics[width=10.5 cm]{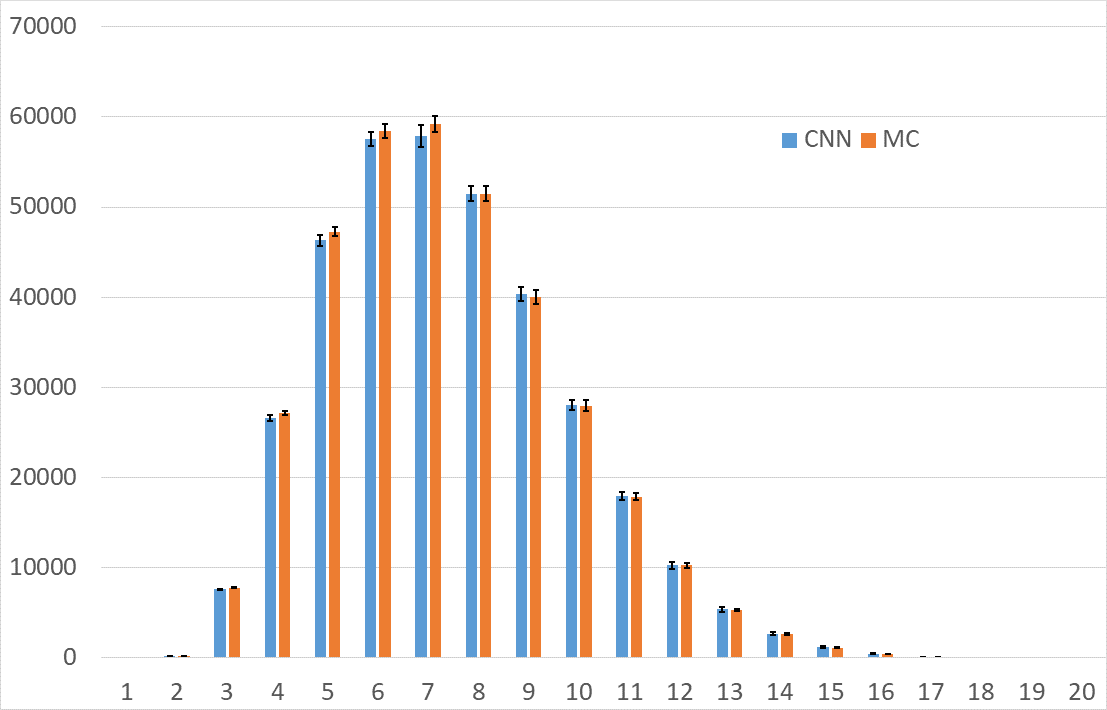}
\caption{Average CNN predictions and the true energy spectrum of 10 runs.
\label{fig_3}}
\end{figure}

\begin{linenomath}
\begin{equation}
    E_{i+1} = E_{i}+0.7\ \text{GeV},\;\;\;E_0 = 0
\end{equation}
\end{linenomath}
for $i$ as the bin number.
The analysis is carried out by associating each event with an array of 20 entries, corresponding to the histograms of $n|_{E_i}^{E^{i+1}}$ with $i=0,1,...,19$.
Employing the $\rho=3$ \textmu m data, we generated 10 random runs (65\% train and 35\% test) and evaluated their histograms.
The average true results, comparing the CNN predictions of the test set are displayed in Fig. \ref{fig_3}.

The CNN predictions have a small average bias over all bins $\left<\Delta N/N\right> = -0.014 \pm 0.016$ and standard deviation $0.020\pm 0.006$, comparing to the true $N$ values.
The average number of positrons per event in these 10 runs was $1024 \pm 31$ for the training and $1000 \pm 46$ for the test data.

To estimate the accuracy of the histogram prediction, we use the conventional KL distance \cite{KL-1951}
\begin{linenomath}
\begin{equation}
    \textrm{KL}(P,Q) = \sum_i{p_i \log(p_i/q_i)}
\end{equation}
\end{linenomath}
\unskip
as the metric for the similarity of two probability distributions $P$ and $Q$. Here
$p_i, q_i$ are respectively the 20 bin heights for these distributions, after their sum is re-normalized to 1.
The KL distance for the case shown in Fig. \ref{fig_3} is $1.2\times 10^{-4}$ indicating that the true distribution isin agreement with the CNN prediction.

\begin{figure}[H]
\includegraphics[width=10.5 cm]{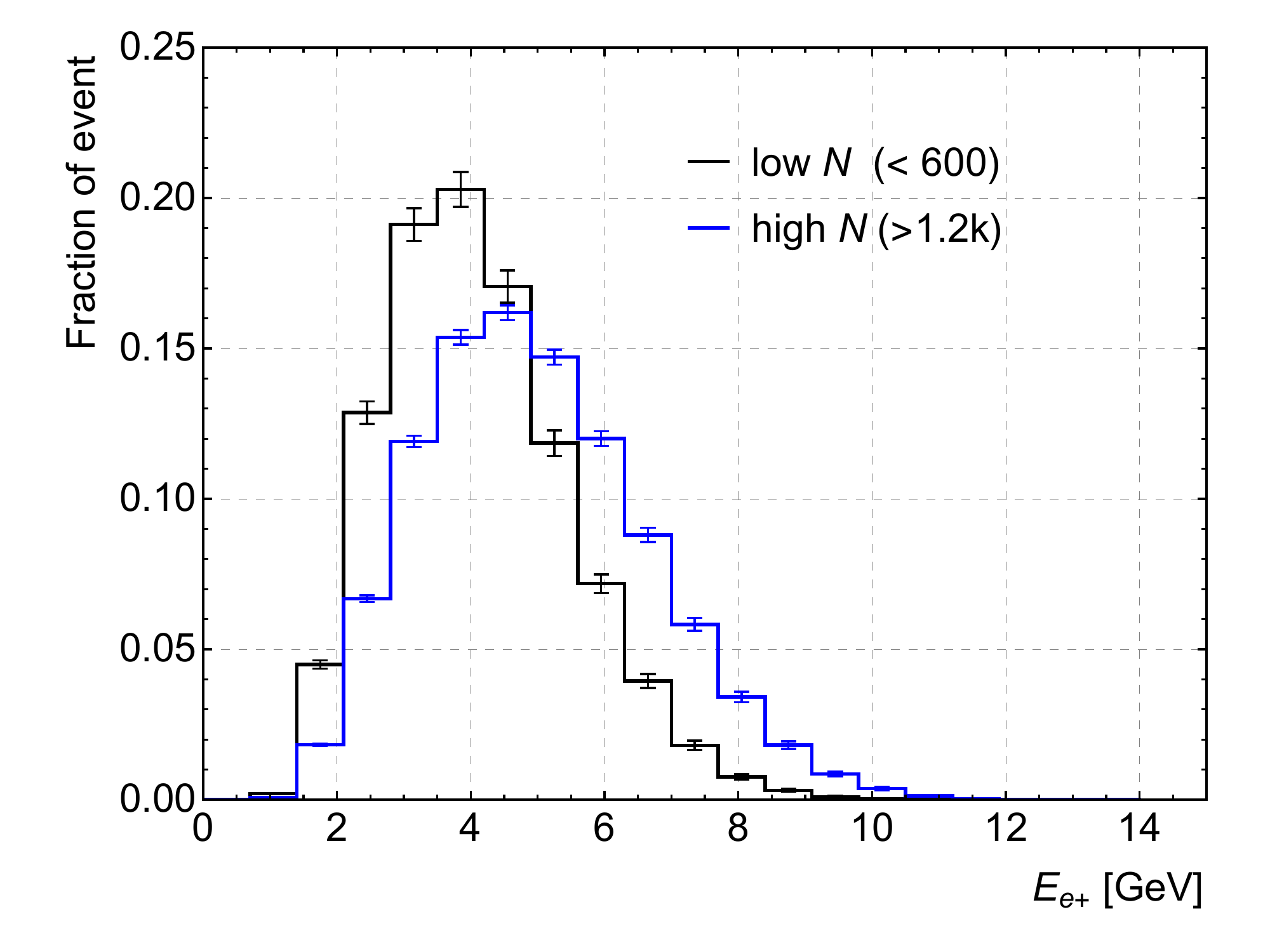}
\caption{Energy histogram of low $N$ events (black) ranges over lower energy bins than the energy histogram of high $N$ events (blue). This is an analysis of the MC data. \label{fig_4}}
\end{figure}

It should be noted that the reconstruction of the test-set distributions by the CNN has been done through accumulating the data event by event.
This means that all events construct harmoniously a coherent distribution, even though the histograms differ from event to event.
As the number of positrons of an event increases, so does its average energy (peak position of the histogram).
This can be seen in Fig. \ref{fig_2}, and is also demonstrated in Fig. \ref{fig_4}, where we plot histograms of the lowest range and of the highest range of $N$ for $\rho=3$ \textmu m events.

Trying to generalize from the $\rho=3$ \textmu m CNN, and applying it to $\rho=5$ \textmu m test data, we have to be careful to adjust the energy ranges accordingly.
When taking care of this aspect, the generalization works very well.
This point is important when we consider future applications to real data, where we have to rely on MC training procedures.

\section{Positron multiplicity and the detector-network system}
\label{sec_5}
Let us turn now to checking the number of positrons hitting the detector, namely the event's multiplicity $N$.
This can be calculated as the summation over all the 20 bins, defined in Eq. (\ref{eq_4}).
Alternatively the CNN can be trained targeting $N$
as demonstrated in Fig. \ref{fig_2}.

An event is recorded by the detector through signals produced by its 24K elements.
This event may be regarded as a point in a 24K-dimensional phase space.
We use the predictability of $N$ as a connection between the CNN representation and the proximity and order properties of integers, in order to establish an intuitive answer.

A few events with the same characteristics, e.g. the same numbers of positrons, occur in the same neighborhood in this large space.
Otherwise, learning the number of positrons by the CNN is impossible: learning implies that all points in this neighborhood be given the same label.
Only then can the procedure be generalized, and the label can be applied to a test-event which appears in the same neighborhood.
The combined detector-network system is somewhat analogous to the visual system, with the detector playing the role of the eye and the network being the analog of the visual cortex.
The success of the CNN lies in correctly identifying the neighborhood which a queried event belongs to.
Predicting N, or predicting the energy distribution of the event, is successful because these labels are attached to the particular neighborhood by the training procedure of the CNN.
The fact that all events with the same labeling end up in the same neighborhood is the important fact.This neighborhood can have multiple regions, as long as these regions are not shared with events of different characteristics.
For this to happen the dynamics should be non-chaotic, i.e. small changes in initial conditions should not lead to large changes of the final region in both the detector phase-space and its representation in the network.
This can also be stated as the condition that clusters of data representing different $N$ values should be well separated within the CNN.

\subsection{Compositionality of CNN prediction}
To demonstrate what happens in our analysis we look for the property of compositionality: given two detector images, $a$ and $b$, will their superposition $a+b$ in the CNN detector-image input lead to $N(a+b)=N(a)+N(b)$?
To perform this test we train the CNN in the regular fashion, and apply it to a test set which is generalized to include such combinations.
The superposition of two images is defined as the sum of all their detector readings, and may be evaluated by the CNN that was trained to predict $N$ of single images.
\begin{figure}[H]
\includegraphics[width=10.5 cm]{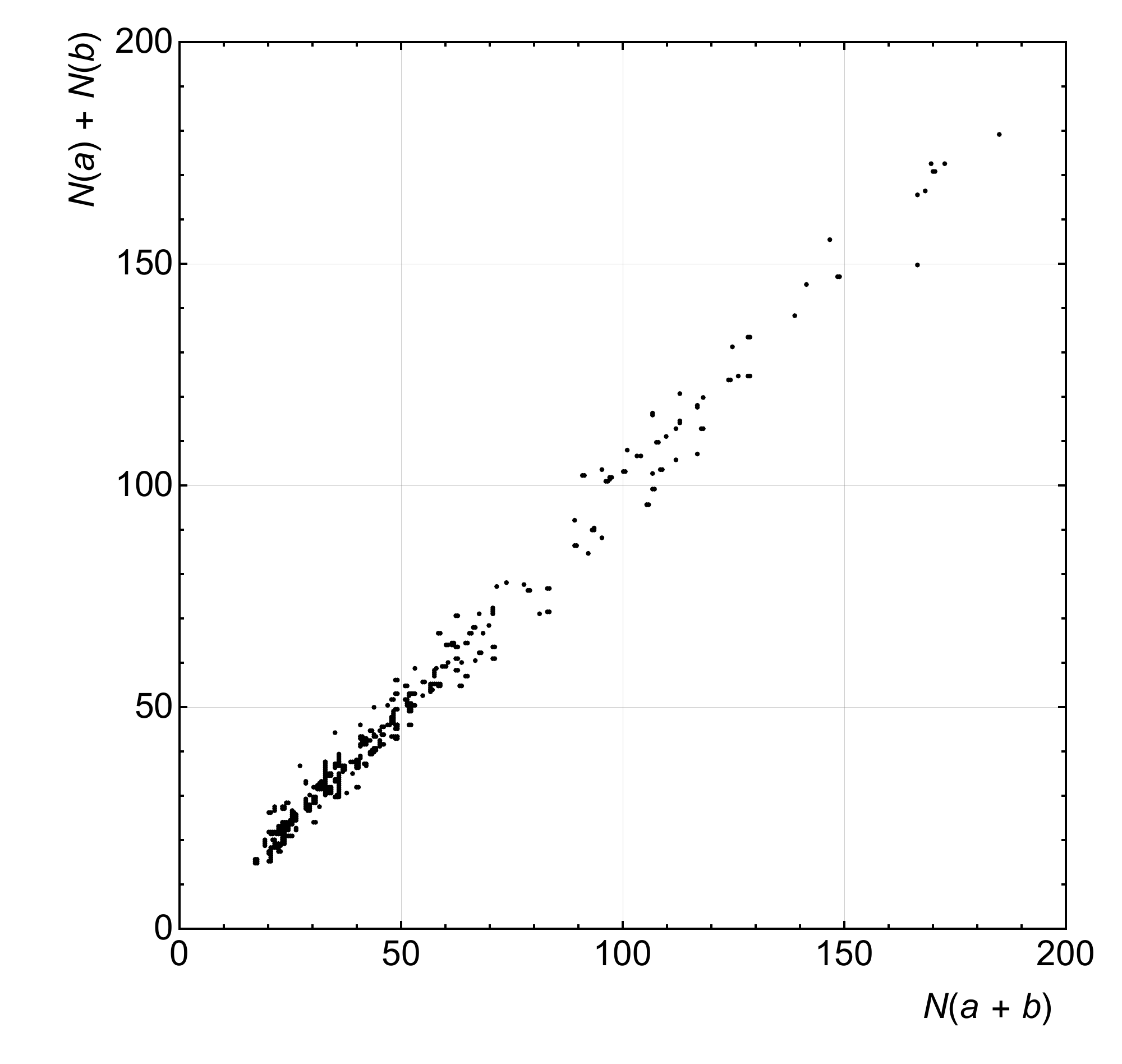}
\caption{Approximate compositionality with linearity along the diagonal. \label{fig_8}}
\end{figure}

We tested this question using the $\rho=5$ \textmu m data.
Intermediate $N$ values of $20<N<200$ show that compositionality holds approximately. The results are displayed in Fig. \ref{fig_8}.
The points in these plots correspond to non-recurrent choices of $(a,b)$ pairs, for which the sum of $N(a)+N(b)$ is plotted vs the network estimate $N(a + b)$.
Although the order of events does not adhere to a strict chain of integers, these data hint that the detector-network system is not very far from it in the shown range of $N$ values.

\subsection{Information reduction}
\begin{figure}[H]
\includegraphics[width=5.2 cm]{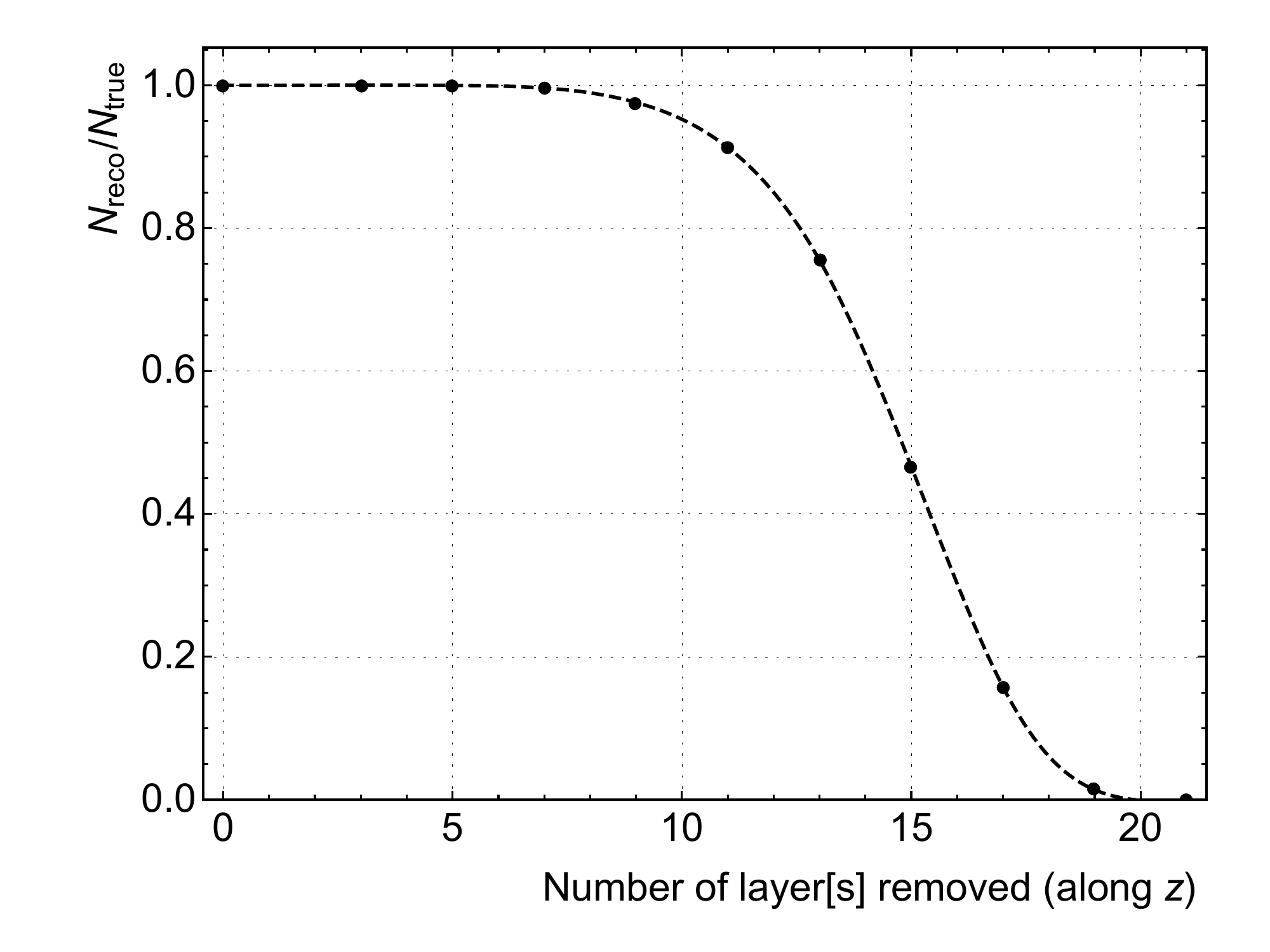}
\includegraphics[width=5.2 cm]{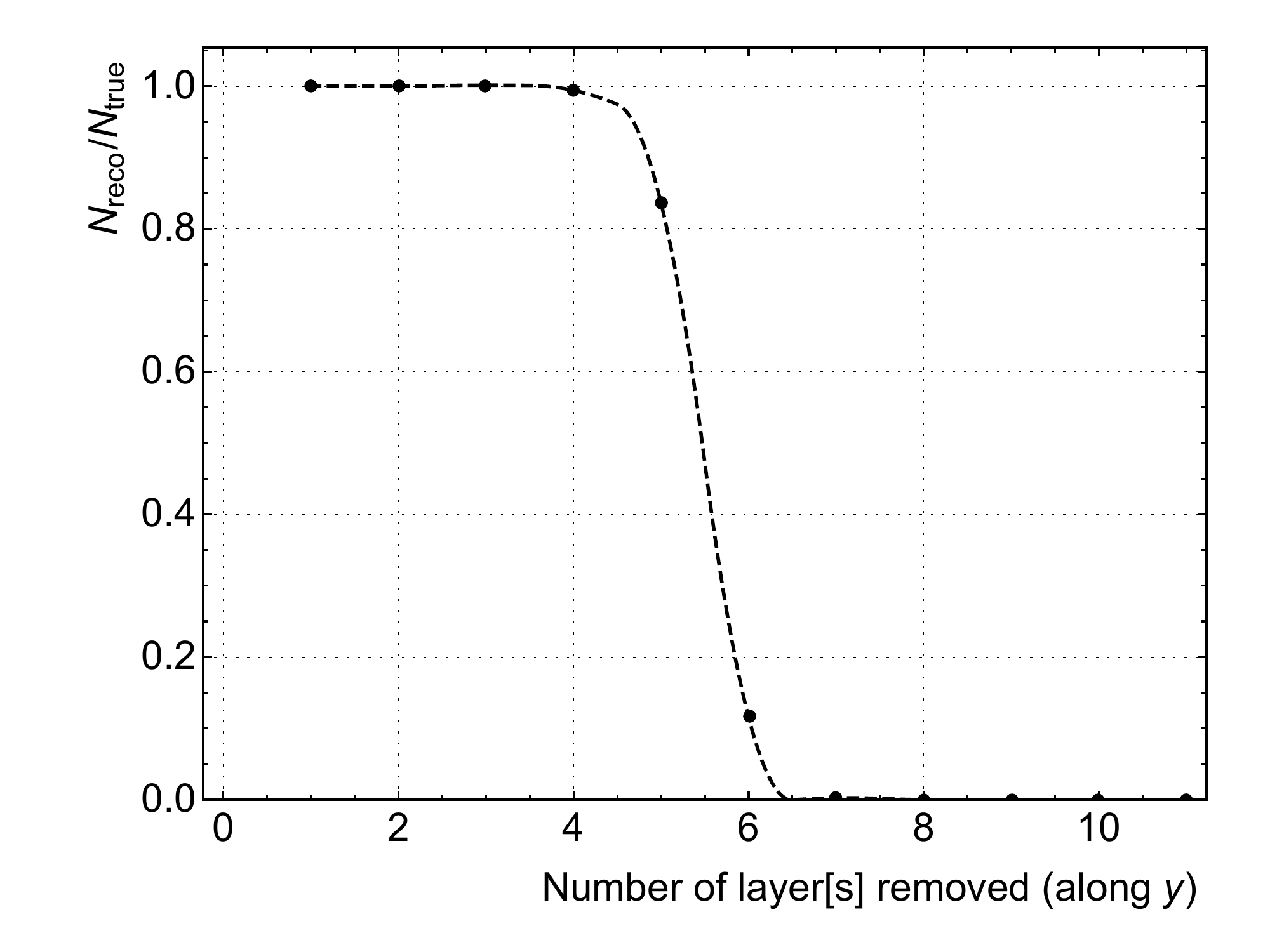}
\caption{Normalized $N$ predictions of test sets for which $z$ (backside forwards) or $y$ (outside inwards) layers have been removed. \label{fig_6}}
\end{figure}

Here we turn to an analysis of the way the events occupy the 24K parameter space of the detector.
Employing the CNN, it is easy to test the information derived from different layers of the detector.
Using our conventional test/train procedure we remove, in the test set only, layers from the detector image in the $z$-direction (starting with its last layer) or in the $y$-direction (starting from the top).
The results for changes in the estimated multiplicity are displayed in Fig. \ref{fig_6}.
In the $z$-direction we find that the last 10 layers (out of 21) have negligible impact, and most of the $N$ information is carried by the first 4--6 $z$-layers.
Similar trends are observed for the prediction of the average energy.
That is to say, the CNN does not employ information from a large number of the detector elements.
This raises the interesting possibility that we may analyze the data by considering a reduced detector image, which we proceed to do by taking into account only three $y$-layers ($y=4,5,6$) and the first 10 $z$-layers.

The resulting energy histograms are very similar, both in shape and in quality, to the ones obtained based on the full detector image. The bias values are again of order 2\%. Note that this is achieved
in a strongly reduced parameter space, using only 3K elements out of the 24K ones of the detector.

\section{Conclusions}
Our study exemplifies the importance of supplying a hardware system, such as an electromagnetic calorimeter in particle physics, with an interpreter software, represented by a deep network.
Whereas the detector has a large number of outgoing signals, the network has a much larger number of parameters which can trivially embed the results of the detector. The non-trivial results of the DL architecture of the ResNet 10 model, are that after training on 50 epochs, this embedding leads to retrievable results on the test sets, allowing us to capture the physical properties of the events which we study.

Moreover, since the network starts out with an image of the detector output, we can easily find out which are the important elements of the detector.
By removing layers from the detector image, we conclude that in the problem at hand we can, using the CNN, retrieve all the physical information we have studied, from an image reduced by a factor of 8.
This is an important conclusion which should be taken into account by the designers of the detector.
Thus the power of CNN software can be harnessed for efficiently designing the hardware needed for experimental technology.

\vspace{6pt}

\funding{This research was partially supported by the Israel Science Foundation and the German Israeli Foundation.}
\acknowledgments{The authors thank Prof. Halina Abramowicz (Tel Aviv University) for her most valuable support and fruitful discussions. The authors also thank the LUXE Collaboration for providing the data of MC simulations.}
\begin{adjustwidth}{-\extralength}{0cm}

\reftitle{References}

\end{adjustwidth}
\end{document}